# Finding knowledge paths among scientific disciplines

Erjia Yan[1]

College of Information Science and Technology, Drexel University, 3141 Chestnut Street, Philadelphia, PA 19104, USA. Phone: (812)606-8091. Fax: (215) 895-2494. Email: erjia.yan@drexel.edu

## Abstract

This paper discovers patterns of knowledge dissemination among scientific disciplines. While the transfer of knowledge is largely unobservable, citations from one discipline to another have been proven to be an effective proxy to study disciplinary knowledge flow. This study constructs a knowledge flow network in that a node represents a Journal Citation Report subject category and a link denotes the citations from one subject category to another. Using the concept of shortest path, several quantitative measurements are proposed and applied to a knowledge flow network. Based on an examination of subject categories in Journal Citation Report, this paper finds that social science domains tend to be more self-contained and thus it is more difficult for knowledge from other domains to flow into them; at the same time, knowledge from science domains, such as biomedicine-, chemistry-, and physics-related domains can access and be accessed by other domains more easily. This paper also finds that social science domains are more disunified than science domains, as three fifths of the knowledge paths from one social science domain to another need at least one science domain to serve as an intermediate. This paper contributes to discussions on disciplinarity and interdisciplinarity by providing empirical analysis.

## Introduction

We live in knowledge societies – societies that are infused by knowledge cultures, epitomized by science and technology, and structured into all walks of social life (Knorr-Cetina, 1999). Knowledge as “intellectual capital” has an immediate impact on the economy, as society has moved from a production-based economy to a knowledge-based one. This shift has significantly changed the organization of labor divisions and has formed new occupations and disciplines (Bell, 1973; Drucker, 1993).

The production and creation of knowledge is not dependent on a single isolated entity; instead, knowledge is diffused, exchanged, and circulated among various entities. Knowledge flow in the past twenty years has become more inter-sectoral, inter-organizational, interdisciplinary, and international (e.g., Lewison, Rippon, & Wooding, 2005; Wagner & Leydesdorff, 2005; Autant-Bernard, Mairesse, & Massard, 2007; Ponds,

[1]Corresponding author

Frenken & Van Oort, 2008; Buter, Noyons, & Van Raan, 2010). Research questions in this area have usually centered on how scientific and technological knowledge, innovative ideas, management skills, or certain influences transfer within different sectors (e.g., Kogut & Zander, 1993; Szulanski, 1996; Storck & Hill, 2000), between different organizations (e.g., Mowery, Oxley, & Silverman, 1996; Narin, Hamilton, & Olivastro, 1997; Cohen, Nelson, & Walsh, 2002; Meyer, 2002), and between different scientific disciplines (e.g., Van Leeuwen & Tijssen, 2000; Rinia, Van Leeuwen, & Bruins, 2001; Rinia et al., 2002; Kiss et al., 2010).

Perceiving knowledge as accumulative and static does not capture the interactive and diversified characteristics of knowledge transfer. As pointed out by Knorr-Cetina (1999), we have a limited understanding of the "contemporary machineries of knowing" (p. 2). One school of thoughts holds that science is disunified – knowledge displays different facades of empirical approaches and different social machines, which brings out the "diversity of epistemic culture" (p. 3). In addition to the trifurcation of sciences (instrumental, practical, and emancipation) in approaches to problem-solving (Habermas, 1988), within each division, various disciplines are expected to have ontological and methodological differences (Suppes, 1984). Meanwhile, other scholars hold the belief that science is becoming unified and claimed that "[t]o further all kinds of scientific synthesis is one of the most important purposes of the unity of science movement" (Neurath , 1996, p. 309). The purpose of this study is to contribute to the discussion by empirically measuring the connectivity of different science and social science disciplines through the application of knowledge paths.

A first attempt in this direction was the study of scientific trading among different disciplines (e.g., Cronin & Davenport, 1989; Cronin & Pearson, 1990; Stigler, 1994; Lockett & McWilliams, 2005; Goldstone & Leydesdorff, 2006; Cronin & Meho, 2008; Larivière, Sugimoto & Cronin, 2012; Yan, Ding, Cronin, & Leydesdorff, 2013). Using the trading metaphor, knowledge exporters and importers were identified in these studies, but they only considered the two ends (exporter and importer) of knowledge flows, without considering the concrete knowledge paths. Consequently, patterns and mechanisms of knowledge flow largely remain in a black box.

Understanding knowledge paths is particularly important to studies of knowledge flows, because it helps us gain insights into patterns of knowledge flows at a more granular level. In addition to questions of what are knowledge importers and exporters, questions such as *how* knowledge is imported and exported can thus be addressed. Specifically, the following questions are investigated in this paper:

- What are knowledge paths among scientific disciplines?

*Previous work*: Previous studies on interdisciplinarity focused mainly on discipline proximity (for clustering and mapping purposes) and not on knowledge flows or knowledge paths;
*Current work*: Using the trading metaphor, knowledge flows among different disciplines are investigated, where the shortest path denotes the most important knowledge flow between two disciplines; and
*Significance*: Patterns of knowledge transfer and dissemination are revealed and empirical analyses are provided;

- What patterns can be found from the identified knowledge paths, and how can these patterns be used to perceive different disciplines?
  *Previous work*: Quantitative studies on the dissemination aspect of disciplinary knowledge were inadequate;
  *Current work*: Quantitative indicators are proposed to measure the extent to which a discipline's knowledge can be accessed by other disciplines, and how distant a discipline's knowledge is to other disciplines; and
  *Significance*: The proposed indicators quantify patterns of knowledge flow and dissemination, providing additional insights into interdisciplinary studies. These indicators are also valuable for scientific evaluation and science policy making.
- How can knowledge be classified, and what additional information can be obtained by evaluating knowledge flows at different knowledge hierarchies?
  *Previous work*: Previous studies usually focused on one type of research entity, such as papers, authors, or journals, but did not systematically study different research entities;
  *Current work*: A six-layered knowledge hierarchy is proposed, allowing one to zoom in to find knowledge paths among papers and also zoom out to gain a holistic view of knowledge transfer between major divisions of knowledge; and
  *Significance*: The proposed knowledge hierarchy is effective in organizing knowledge, and can be easily adopted in studies of clustering and mapping research specialties.
- What is the backbone of knowledge flow in the knowledge flow network?
  *Previous work*: Most previous science maps are similarity-based (i.e., co-citation, bibliographic coupling, co-word). Weighted, directed knowledge flow maps were not well explored;
  *Current work*: In addition to the quantitative indicators, the backbone of knowledge flow is visualized.; and
  *Significance*: Knowledge flow maps are informative to scientists, scholars, science policy makers, and the general public and help the understanding of how knowledge is disseminated among disciplines.

## Literature review

## Inter-sectoral and inter-organizational knowledge flows

Efforts in inter-sectoral and inter-organizational knowledge flows are largely employed to understand the mechanisms of different types of knowledge transfers and to identify ways to facilitate knowledge transfers between organizations. Cohen, Nelson and Walsh (2002) discussed the different forms of knowledge flows in science and technology, including how knowledge may be transferred by personal contacts at conferences and workshops, by mobility of researchers, by advisor-advisee relationships, by collaborative research projects, or by publication channels such as scientific articles and patents. Using survey data, the authors found that R&D managers in U.S. firms considered publications to be the dominant channel of knowledge flow. Zellner and Fornahl (2002) made a similar argument, positing that the major organizational knowledge acquisition channels are the recruitment of people, the information networks of employees, and the formal cooperation networks. Szulanski (1996) exemplified that the major barriers of inter-sectoral knowledge transfer lie in knowledge recipients' lack of absorptive capacity, causal ambiguity, and a tension between knowledge senders and knowledge recipients. These findings were confirmed by Almeida and Kogut (1999), who found that the mobility of employees influences the organizational knowledge transfer, and such knowledge transfer is in turn embedded in the information networks of employees.

Similar to many important concepts in economics, the transfer of knowledge is largely unobservable (Jaffe, Trajtenberg, & Fogarty, 2000) and thus requires proxies to measure concepts of interest. Patent citations provide a practical instrument to quantitatively study knowledge flow in empirical work. In a pioneering research of Jaffe, Trajtenberg and Henderson (1993), the authors studied the association between geographical locations and citation intensity. They found that domestic patents are more likely to cite other domestic patents, yet these localization patterns slowly became less significant in later years. In the same vein of research, Jaffe and Trajtenberg (1999) found the transfer of knowledge in patent citation networks is restrained by country boundaries as well as organizational boundaries and patent classes. MacGarvie (2005) used the notion of proximity to interpret such phenomenon, in that knowledge diffusion is enhanced by physical and technological proximity. In addition to studies of organizational boundaries in knowledge flow, efforts have also compared knowledge diffusion decay for different types of organizations in patent citation networks. For instance, Bacchiocchi and Montobbio (2007) found that the knowledge embedded in university and public research patents tends to diffuse more rapidly than corporate patents. Besides citation data, empirical research has also used coauthorship and survey data to study inter-sectoral and inter-organizational knowledge flow (e.g., Cohen, Nelson, & Walsh, 2002; Meyer, 2002).

Predicated upon previous endeavors in social network analysis, we have witnessed a growing number of studies in the past decade that used network-based approaches to examine knowledge flow in organizations. These studies paid particular attention to the structural significance of organizations in knowledge flow networks. For example, Tsai (2001) analyzed the network positions of organizations and found that network positions have significant and positive effects on business innovation and performance. Studies have also shown that social cohesion and network range (Reagans & McEvily, 2003), social capital (Inkpen & Tsang, 2005), and betweenness (Levin & Cross, 2004) are contributing factors that facilitate knowledge transfer in organizations.

**Interdisciplinary knowledge flows**

The quantitative studies of knowledge flow among disciplines have usually used citations as the research instrument. Citations between scientific articles imply a knowledge flow from the cited entity to the citing entity (Jaffe, Trajtenberg, & Henderson, 1993; Van Leeuwen & Tijssen, 2000; Nomaler & Verspagen, 2008; Mehta, Rysman, & Simcoe, 2010). As citations can be aggregated into several levels, such as paper/patent-, author-, journal-, institution-, and discipline-level, scholars have investigated knowledge flows in patent citation networks (e.g., Jaffe, Trajtenberg, & Henderson, 1993; Narin, Hamilton, & Olivastro, 1997; Jaffe, Trajtenberg, & Fogarty, 2000; Chen & Hicks, 2004; Mehta, Rysman, & Simcoe, 2010), paper citation networks (e.g., Chen, Zhang, Vogeley, 2009; Shi, Tseng, & Adamic, 2009), author citation networks (e.g., Zhuge, 2006), journal citation networks (e.g., Alvarez & Pulgarín, 1997; Frandsen, Rousseau, & Rowlands, 2006), institution citation networks (e.g., Börner, Penumarthy, Meiss, & Ke, 2006; Yan & Sugimoto, 2011), and discipline citation networks (e.g., Van Leeuwen & Tijssen, 2000; Rinia, Van Leeuwen, & Bruins, 2001; Rinia et al., 2002).

In a pioneering quantitative investigation of interdisciplinary dependency, Borgman and Rice (1992) studied cross-disciplinary citations between information science and communication science journals, and found that a few information science journals heavily cited communication science journals. Leydesdorff and Probst (2009) also found that communication science journals were cited frequently by political science and social psychology journals. Using a 30-year citation data set on information science publications, Cronin and Meho (2008) found that information science has become a more successful exporter of knowledge that is less introverted, as more recent articles in information science have cited and are being cited more intensively by articles in fields such as computer science, engineering, business and management, and education. The findings were confirmed by Levitt, Thelwall, and Oppenheim (2011), where the authors found that library and information science had the largest increase in interdisciplinarity between 1990 and 2000 among different social science fields.

Scholars have also used various statistical models, such as the epidemiological model (Bettencourt et al., 2006; Kiss et al., 2010), the population contagion model (Bettencourt et al., 2008), the clique percolation method (Herrera, Roberts, & Gulbahce, 2010), the small-world model (Cowan & Jonard, 2004), and the diffusion model (Zhuang, Chen, &

Feng, 2011) to describe and simulate the creation and dissemination of knowledge in different contexts.

Epidemiological models focus on the transmission of different traits among certain populations; such traits can be transmitted diseases, behaviors, or innovative ideas. An individual can be classified into one of the basic classes in epidemiological models: the susceptible (S) class, the exposed (E) class, the infected class (I), the skeptical class (Z), and the recovered (R) class (Hethcote, 2000). While scholars can choose these classes based on their specific research questions, there is always a trade-off between the level of detail and the complexity of the model, as pointed out by Bettencourt et al. (2006), who compared several epidemiological models with the goal to explore the spread of the Feynman diagram in post-war U.S., U.S.S.R., and Japan, including SIR, SIZ, SEI, and SEIZ models. The authors found that all four models have an accurate fit with empirical data, and among them, SEIZ captured most adequately the role of different classes in the transmission process and yielded the best fits. Kiss et al. (2010) used an epidemiological model to describe the spread of research topics across disciplines. They conducted a case study on *kinesin* research publications and found that the diffusion of topics is more likely to occur between disciplines with existing knowledge flows, and this diffusion across disciplines takes a considerable amount of time (4.0 to 15.5 years).

Previous endeavors on inter-sectoral, inter-organizational, and inter-disciplinary knowledge transfer laid sound theoretical and methodological foundations for inquiries on knowledge flows. Based on our best knowledge, however, there has been no study to date on finding knowledge paths among scientific disciplines. This study attempts to fill this gap by exploring patterns of knowledge flows at several knowledge hierarchies, including subjects, classes, and top divisions of knowledge.

## Methods

### The construction of knowledge flow networks

Subject category citation data from ISI (Thomson Reuters) Journal Citation Report were used. Despite discussions on the accuracy of subject categories (e.g., Boyack, Klavans, & Börner, 2005; Rafols & Leydesdorff, 2009), the ISI classification scheme has been widely used and accessed (Zitt, 2005; Van Raan, 2008) (for procedures on data collection, please refer to Rafols and Leydesdorff, 2009). It is thus used as an empirical proxy to study disciplinary knowledge flows. The limitation of using ISI data is deliberated on in the *Discussion* section. A knowledge flow network, or a field-to-field citation network, is constructed based on the 2009 data set (in this paper, the terms of discipline, field, and domain are used interchangeably). Citations from a multi-assigned journal are counted towards all assigned subject categories; this multi-assignment is also referred to as "multiple counting" (Yan, Ding, Cronin, & Leydesdorff, 2013). "Multiple counting" avoids the arbitrariness of assigning a multi-assigned journal to any one subject category; however, the caveat is that such assignment may result in citation inflation (e.g., *Journal of the American Society for Information Science and Technology* is assigned to two categories: Information Science & Library Science and Computer Science Information Systems; thus, its citation flow is counted toward both categories). Figure 1(a) shows an example of a knowledge flow network consisting of five fields.

| a | | Cited Field | | | | |
|---|---|---|---|---|---|---|
| | | A | B | C | D | E |
| Citing Field | A | 200 | 100 | 20 | 70 | 300 |
| | B | 20 | 450 | 220 | 60 | 10 |
| | C | 320 | 50 | 550 | 100 | 20 |
| | D | 40 | 20 | 80 | 325 | 10 |
| | E | 50 | 250 | 40 | 90 | 500 |

| b | | Cited Field | | | | |
|---|---|---|---|---|---|---|
| | | A | B | C | D | E |
| Citing Field | A | 5.E-03 | 1.E-02 | 5.E-02 | 1.E-02 | 3.E-03 |
| | B | 5.E-02 | 2.E-03 | 5.E-03 | 2.E-02 | 1.E-01 |
| | C | 3.E-03 | 2.E-02 | 2.E-03 | 1.E-02 | 5.E-02 |
| | D | 3.E-02 | 5.E-02 | 1.E-02 | 3.E-03 | 1.E-01 |
| | E | 2.E-02 | 4.E-03 | 3.E-02 | 1.E-02 | 2.E-03 |

Figure 1. An example of a knowledge flow network

**Measurement**

In order to characterize the knowledge flow among fields, the Dijkstra algorithm (Dijkstra, 1959) was used to search for shortest paths between any two fields in the knowledge flow network. The idea of the Dijkstra algorithm is to find a path between two nodes so that the sum of edge weights reaches the minimum in a weighted directed network.

Since the shortest path is defined in a distance-based network, the flow distance between fields *i* and *j* is operationalized by *reverse_flow_width* (Figure 1(b)): the more citations from one field to another, the wider the knowledge flow, and thus the shorter the flow distance:

$$flow_dist_{i \to j} \stackrel{\text{def}}{=} reverse_flow_width_{i \to j} = \frac{1}{number\ of\ citations\ from\ j\ to\ i} \quad (1)$$

The initiation of the algorithm involves setting the weight of all pairs of knowledge paths as infinite: $path_weight_{i \to j} = inf, for\ i, j = 1{:}n$, and then setting the weight of knowledge paths to oneself as zeros: $path_weight_{i \to i} = 0, for\ i = 1{:}n$. The major step is to update the path weight with network information. For instance, if the existing weight for $path_weight_{i \to j}$ is larger than new weight $path_weight_{i \to k} + flow_dist_{k \to j}$, then the existing weight is updated by the new weight:

$path_weight_{i \to j} = path_weight_{i \to k} + flow_dist_{k \to j}$, in this way until all pairs are traversed.

In Figure 2, two examples (shortest knowledge path from A to E and shortest knowledge path from E to A) are given based on the sample knowledge flow network illustrated in Figure 1. In Figure 2, the knowledge path from Field A to Field E follows A->C->B->E, and the knowledge path from Field E to Field A follows E->A.

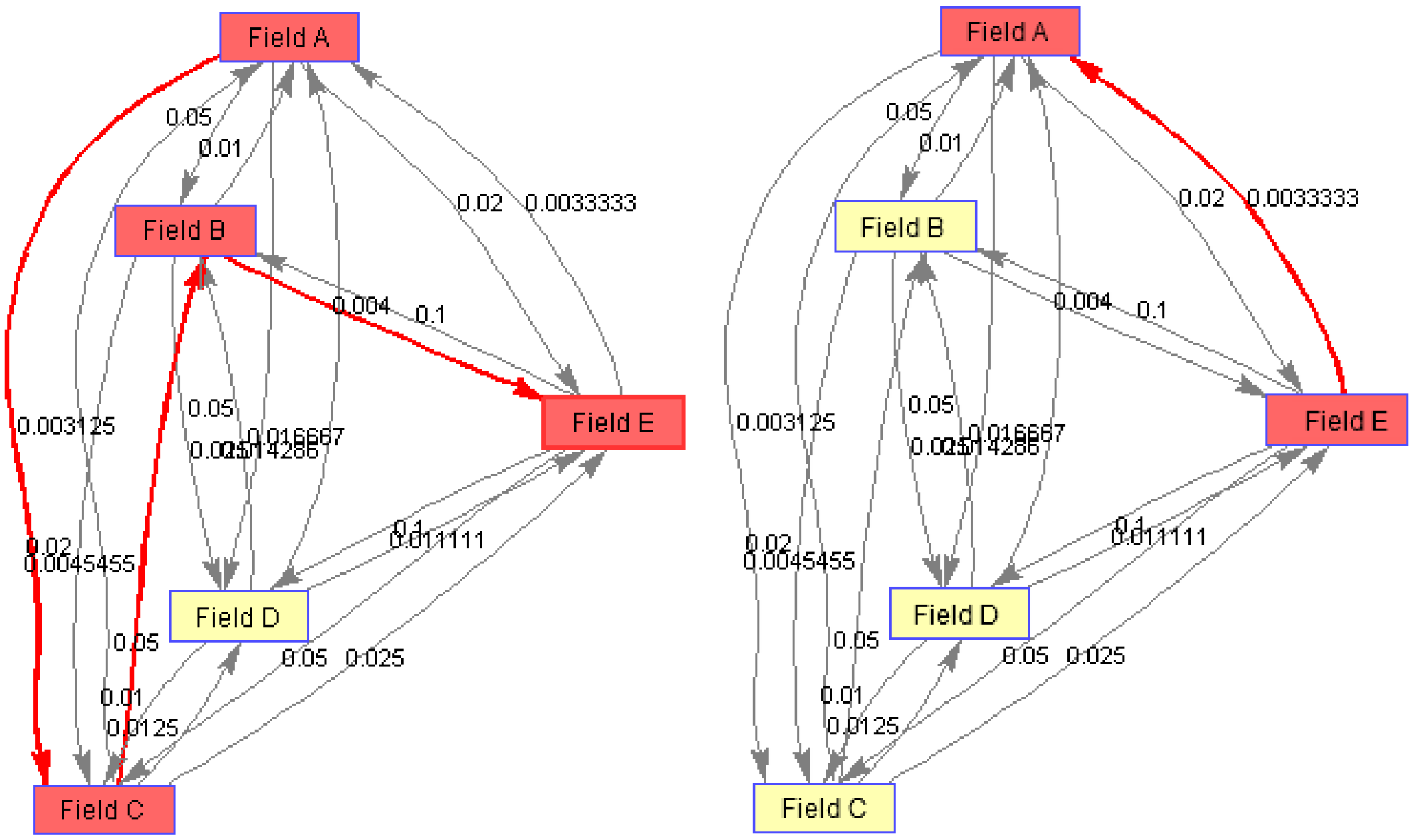

Figure 2. Examples of finding shortest knowledge paths between fields

Note that the shortest path algorithm is guided by the assumption that a citation flow denotes a knowledge flow. Although this assumption is supported by previous studies (e.g., Borgman & Rice, 1992; Wouters, 1998; Cronin & Meho, 2008; Yan, Ding, Cronin, & Leydesdorff, 2013), it may simplify patterns of knowledge dissemination. In reality, disciplinary knowledge may absorb and incorporate new body of knowledge before it flows to other fields, as Nerkar (2003) argued that "…consider new knowledge creation as a recombinant process that involves search, discovery, and use of existing, codified, and observable knowledge" (p. 213).

## Aggregation levels

Currently, two classification schemes are widely recognized and used in studies of scholarly communications: the Essential Science Indicator (ESI) classes and the Journal Citation Report (JCR) subject categories. Both are maintained by Thomson Reuters. In ESI, journals are grouped into 22 classes, and in JCR, journals are grouped into more than 200 subject categories. Formally, ESI and JCR are two separate classification schemes. Based on their journal assignment, in this study, JCR subject categories are mapped into the ESI classes[2]. For example, the ESI class *Physics* contains nine JCR subject categories: (1) *Acoustics*, (2) *Optics*, (3) *Physics, Applied*, (4) *Physics, Condensed Matter*, (5) *Physics, Fluids & Plasmas*, (6) *Physics, Mathematical*, (7) *Physics, Multidisciplinary*, (8) *Physics, Nuclear*, and (9) *Physics, Particles & Fields*. In the proposed knowledge hierarchy, knowledge is thus divided into sciences, social sciences, and arts and humanities, and these three comprise 22 ESI classes that can be further divided into more than 200 JCR subject categories. Each subject category contains certain numbers of journals, from several dozen to a few hundred, and each journal publishes articles periodically (Figure 3). This study focuses on the analysis of the three middle knowledge hierarchies: sciences/social sciences, ESI classes, and JCR subject categories.

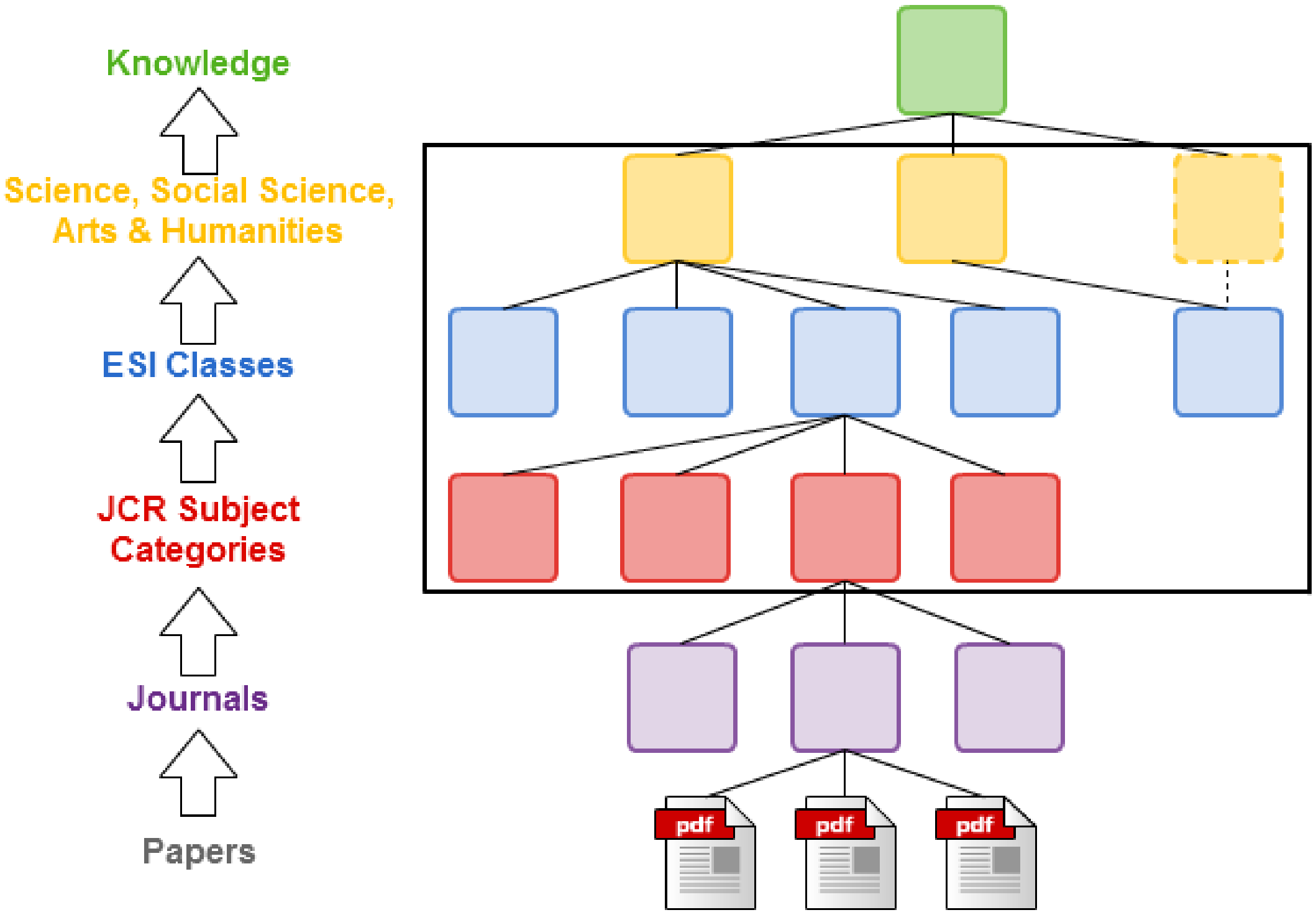

Figure 3. Knowledge hierarchy (Arts & Humanities is displayed in the dotted line as it is not included in the current study)

## Evaluative indicators

For an effective analysis, several indicators are proposed, including average shortest path length, average shortest path weight, and occurrence in shortest path. The weighted directed network can be represented as $G=(V, A)$, where $A$ represents the weighted directed link set and *V* represents the vertex set of fields. The proposed indicators are formally defined as:

- Shortest path (SP) from *i* to *j* ($SP_{i \to j}$) is a path from *i* to *j* in the knowledge flow network such that the sum of the distances of its constituent links is minimized, where the distance is defined in formula (1);
- Shortest path length (SPL) from *i* to *j* ($SPL_{i \to j}$) is defined as the number of nodes traversed in transferring a piece of information in the shortest path from *i* to *j* ($SP_{i \to j}$);
- Average shortest path length (ASPL) for *i* as the source of knowledge transfer is defined as $ASPL_{i:source} = \frac{\sum_{j=1}^{n} SPL_{i \to j}}{n}$, where *n* is the number of subject categories in this study;
- ASPL for *i* as destination of knowledge transfer is defined as $ASPL_{i:destination} = \frac{\sum_{j=1}^{n} SPL_{j \to i}}{n}$;
- Shortest path weight (SPW) from *i* to *j* ($SPW_{i \to j}$) is defined as the accumulative link weights in the shortest path from *i* to *j* ($SP_{i \to j}$), where the weight is defined in formula (1);
- Average shortest path weight (ASPW) for *i* as the source of knowledge transfer is defined as $ASPW_{i:source} = \frac{\sum_{j=1}^{n} SPW_{i \to j}}{n}$;
- ASPW for *i* as the destination of knowledge transfer is defined as $ASPW_{i:destination} = \frac{\sum_{j=1}^{n} SPW_{j \to i}}{n}$; and
- Occurrence in shortest path (OiSP) for *k* is defined as the number of times *k* occurred in shortest paths between all pair of nodes: $\sum_{i=1}^{n} \sum_{j=1}^{n} (k \text{ is on the shortest path of } SP_{i \to j}?, 1, 0)$.

[2] The matching results can be found at http://ella.slis.indiana.edu/~eyan/papers/field/ESI-SC.txt

As the source of a knowledge flow, ASPL denotes how easily its knowledge can be accessed by other fields (or the inclination of other fields to access its knowledge). As to the destination of a knowledge flow, ASPL denotes how easily it can access other's knowledge (or the inclination of the target field to access the knowledge of other fields). The average shortest path weight (ASPW) measures how distant or how different a field's knowledge is to other fields (as the source of knowledge flow) or from other fields (as the destination of knowledge flow). The occurrence in the shortest path (OiSP) denotes how important a field is to other fields' knowledge transfer, and is an indicator related to betweenness centrality. The standard form of betweenness centrality, however, can only be applied to unweighted and undirected networks. As link weights and directions are crucial in studying knowledge flows, the number of occurrences is used to measure the role in which each field functions in connecting different sources of knowledge. A field with a higher occurrence in the shortest path thus plays a more important role in connecting various knowledge sources.

The evaluative indicators are first applied to the field-to-field knowledge flow network, and then the results are aggregated into the class and top division levels, as shown in Figure 4.

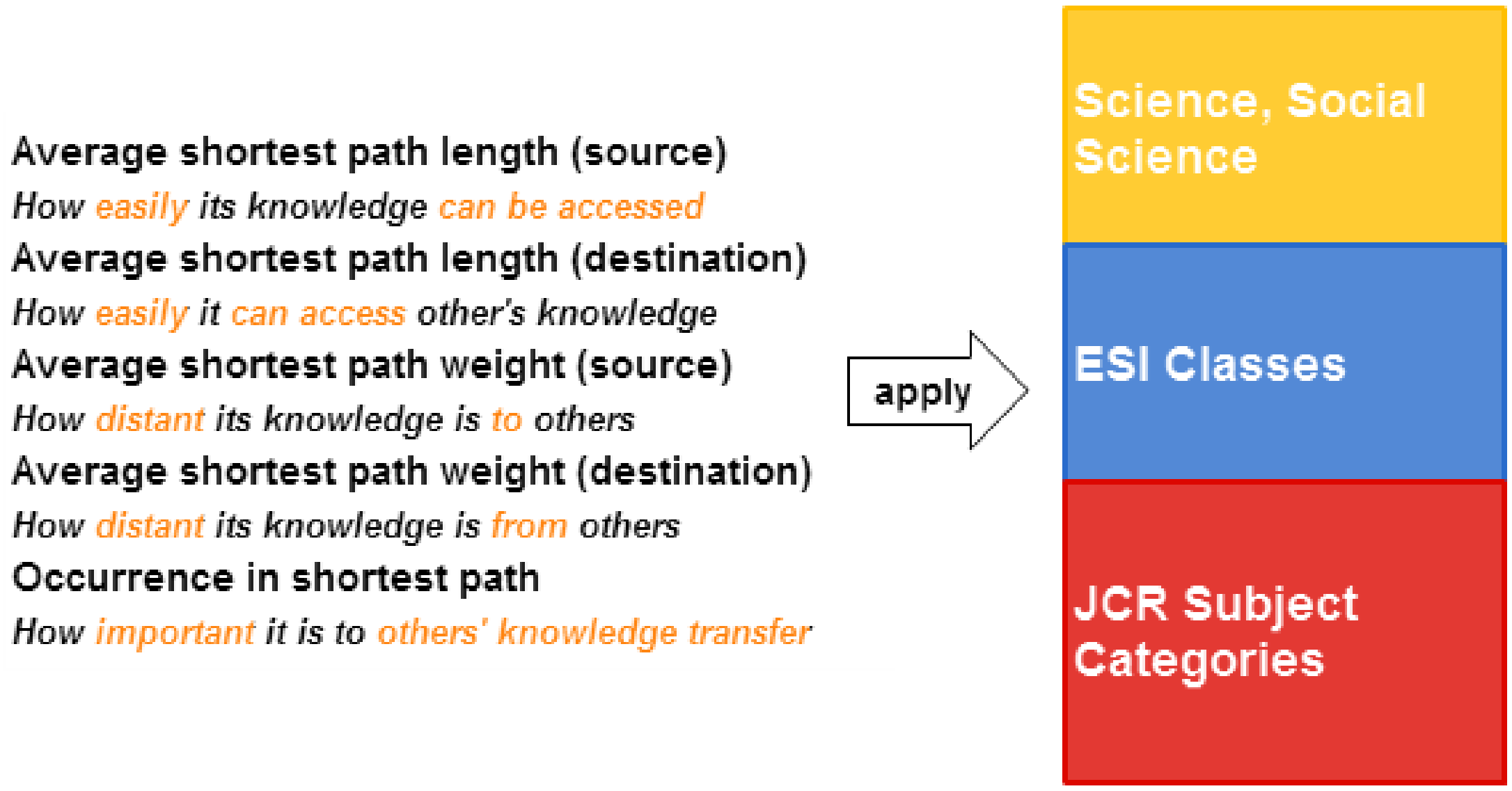

Figure 4. Proposed indicators to measure knowledge flows

## Results

### Results on subject categories

This section first introduces results on subject categories, and then moves to results on ESI classes and top divisions of knowledge. Knowledge paths are identified for each pair of subject categories. In total, there are 48,841 knowledge paths among 221 subject categories. The shortest knowledge path one is from a subject category to itself. The length of the longest knowledge path is 14, suggesting that as many as 14 subject categories are involved in transferring a piece of knowledge between two subject categories. The distribution does not pass the normal significance test with Kolmogorov-Smirnov's asymptotic significance equal to 0 ($p<0.05$). The distribution is positively skewed, suggesting that there are outliers whose shortest paths are much longer than the median of 6.

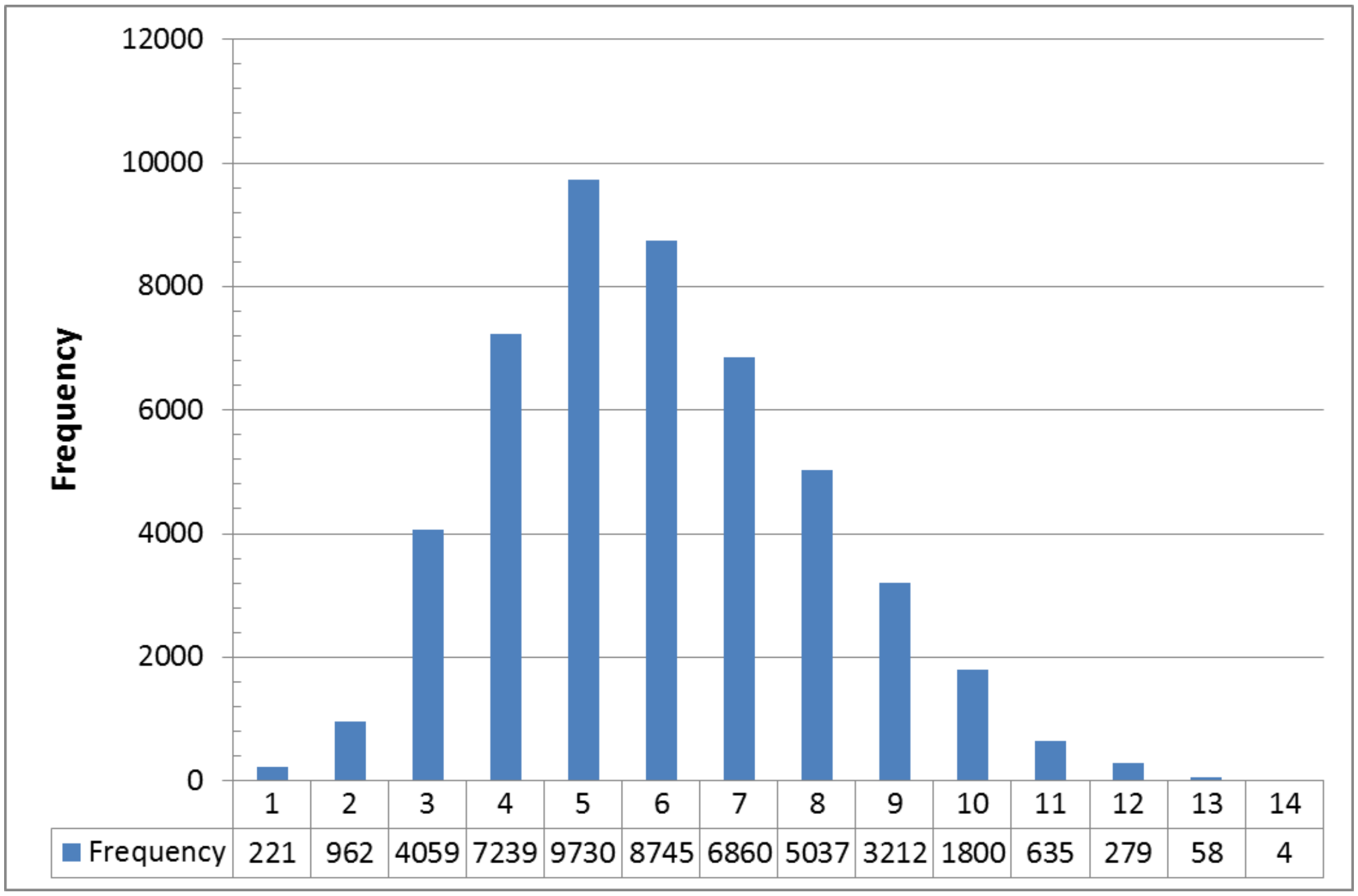

Figure 5. Distribution of knowledge path lengths

For any discipline, a longer average shortest path length (as the source of knowledge flow) suggests its knowledge flow process is more difficult than that of other disciplines; a longer average shortest path length (as the destination of knowledge flow) suggests it would be more difficult for other disciplines to export knowledge to the target discipline. Horizontally, each cell in Figure 6 denotes how easily a discipline's knowledge can be accessed by other disciplines; vertically, each cell denotes how easily a discipline can access other disciplines' knowledge, in that the legend shows the path length (from 1 to 14).

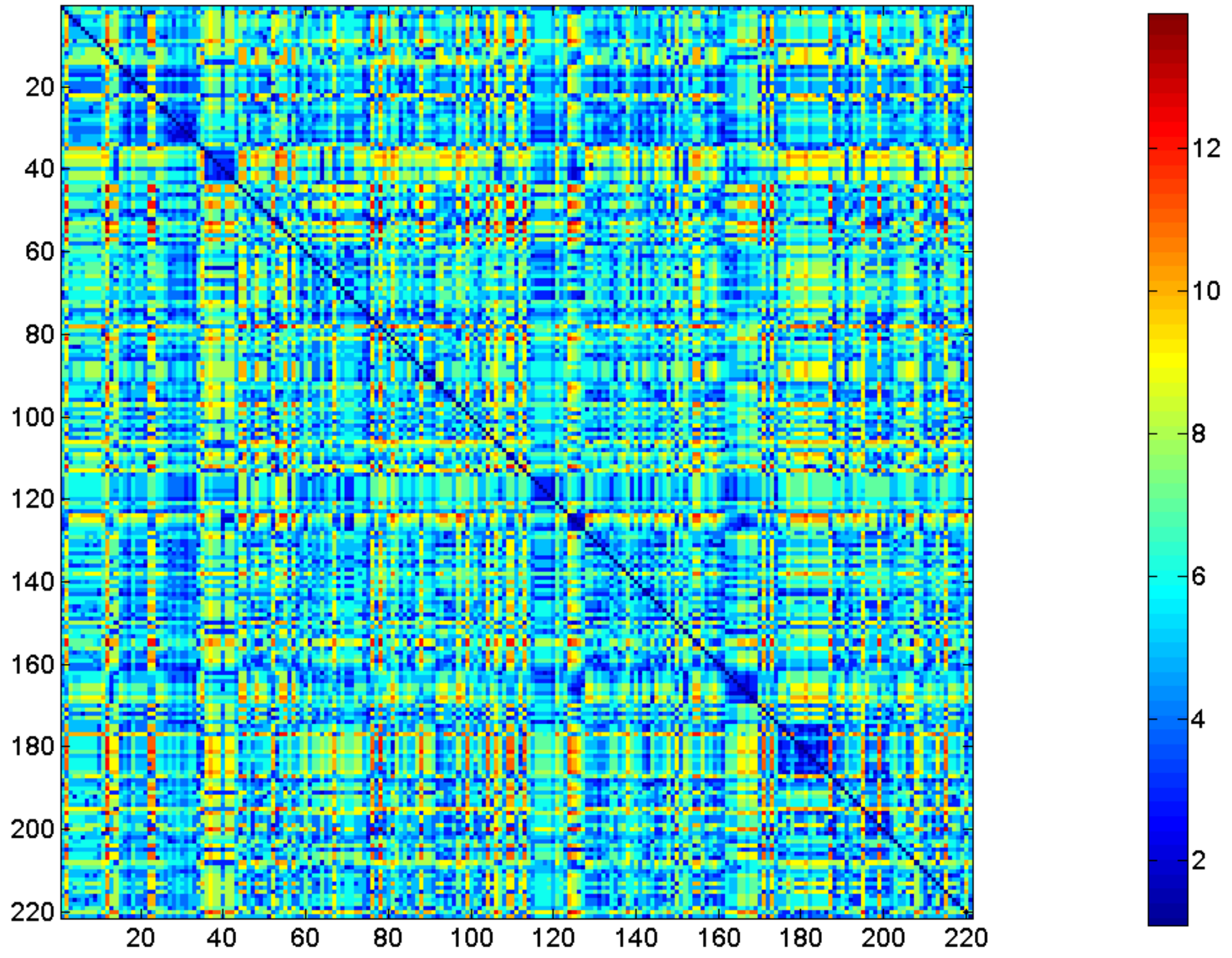

Figure 6. Heat map on shortest path length for JCR subject categories

Discrepancies can be found, in that some disciplines are more reluctant to import other disciplines' knowledge (having more "red" cells in a column per Figure 6) and their own knowledge is more difficult to be accessed by other disciplines (having more "red" cells in a row), while knowledge can flow in and out from some disciplines more easily (having more "blue" cells in columns and rows, respectively).

For the top ten science and social science disciplines listed in Table 1, their average shortest path lengths (as the source of knowledge flow) are shorter, suggesting that their domain knowledge can flow to other disciplines more easily. The top ten science disciplines are mainly composed of biomedical related fields; comparatively, the top ten social science disciplines are more diversified. By way of contrast, the average shortest path lengths for the top ten social science disciplines are longer than those for the top science disciplines.

Table 1. Top ten science and social science subject categories based on average shortest path length (as the source of knowledge flow)

| Subject categories (Science) | Average shortest path length | SD | Max | Subject categories (Social Science) | Average shortest path length | SD | Max |
|---|---|---|---|---|---|---|---|
| Biochemistry & Molecular | 4.19 | 1.71 | 9 | Sociology | 4.72 | 1.53 | 9 |
| Biology | | | | | | | |
| Chemistry, Multidisciplinary | 4.42 | 1.37 | 8 | Economics | 5.06 | 1.94 | 9 |
| Statistics & Probability | 4.48 | 1.28 | 7 | Social Sciences, Interdisciplinary | 5.28 | 1.75 | 10 |
| Public, Environmental & Occupational Health | 4.49 | 1.75 | 9 | Demography | 5.33 | 1.74 | 10 |
| Chemistry, Physical | 4.55 | 1.32 | 7 | Hospitality, Leisure, Sport & Tourism | 5.34 | 1.40 | 8 |
| Pharmacology & Pharmacy | 4.58 | 1.53 | 9 | Environmental Studies | 5.37 | 1.79 | 9 |
| Computer Science, Interdisciplinary Applications | 4.61 | 1.30 | 8 | Ergonomics | 5.47 | 1.79 | 10 |
| Mathematical & Computational Biology | 4.62 | 1.30 | 9 | Transportation | 5.47 | 1.79 | 10 |
| Biology | 4.69 | 1.33 | 9 | Health Policy & Services | 5.48 | 1.77 | 10 |
| Materials Science, Multidisciplinary | 4.70 | 1.40 | 8 | Social Sciences, Biomedical | 5.48 | 1.77 | 10 |

Besides the measurement of shortest path length, shortest path weight is another useful indicator to study patterns of knowledge flows in scientific disciplines. The measurement of shortest path length focuses on the “steps” it takes to transfer knowledge, and the shortest knowledge path weight focuses on the “citation distance” of a discipline to another discipline.

The top ten science and social science disciplines based on average shortest path weight (as source of knowledge flow) is shown in Table 2.

Table 2. Top ten science and social science subject categories based on average shortest path weight (as source of knowledge flow)

| Subject categories (Science) | Average shortest path weight | Subject categories (Social Science) | Average shortest path weight |
|---|---|---|---|
| Biochemistry & Molecular Biology | 2.22E-04 | Psychology, Clinical | 2.72E-04 |
| Cell Biology | 2.25E-04 | Psychology, Experimental | 2.97E-04 |
| Chemistry, Multidisciplinary | 2.27E-04 | Rehabilitation | 3.30E-04 |
| Neurosciences | 2.28E-04 | Psychology, Multidisciplinary | 3.31E-04 |
| Genetics & Heredity | 2.28E-04 | Psychology, Developmental | 3.46E-04 |
| Chemistry, Physical | 2.29E-04 | Social Sciences, Biomedical | 3.48E-04 |
| Materials Science, Multidisciplinary | 2.30E-04 | Economics | 3.74E-04 |
| Oncology | 2.30E-04 | Psychology, Social | 3.76E-04 |
| Biophysics | 2.30E-04 | Business, Finance | 4.15E-04 |
| Pharmacology & Pharmacy | 2.30E-04 | Management | 4.34E-04 |

The top ten science disciplines are dominated by biomedical-related disciplines and the top ten social science disciplines are dominated by psychology-, economics-, and business- related disciplines. Knowledge from these disciplines is thus more closely related to other disciplines, which may imply a permeable and/or interdisciplinary character.

As shown in Table 3, among science disciplines, biochemistry, chemistry, and materials science have a higher occurrence in shortest paths and are thus the most important disciplines in brokering scientific knowledge; among social science disciplines, economics and psychology have a higher occurrence in shortest paths and are thus the most important disciplines in interconnecting social science knowledge.

Table 3. Top ten science and social science subject categories based on occurrence in shortest path

| Subject categories (Science) | Occurrence in shortest path | Subject categories (Social Science) | Occurrence in shortest path |
|---|---|---|---|
| Biochemistry & Molecular Biology | 24,805 | Economics | 6,375 |
| Chemistry, Multidisciplinary | 16,756 | Psychology, Clinical | 1,826 |
| Materials Science, Multidisciplinary | 16,034 | Psychology, Social | 1,312 |
| Neurosciences | 13,236 | Sociology | 1,295 |
| Environmental Sciences | 12,832 | Psychology, Developmental | 1,169 |
| Chemistry, Physical | 10,208 | Business | 1,143 |
| Physics, Applied | 6,848 | Political Science | 1,098 |
| Physics, Multidisciplinary | 6,394 | Rehabilitation | 1,094 |
| Physics, Condensed Matter | 6,292 | Management | 890 |
| Engineering, Electrical & Electronic | 6,156 | Psychology, Experimental | 884 |

### Results on ESI classes

Similar to heat maps used in the preceding section, horizontally, each cell in Figure 7 denotes how easily an ESI class's knowledge can be accessed by other classes; vertically, each cell denotes how easily a class can access other classes' knowledge. Names for the 21 ESI classes can be found in Table 4 (the Multidisciplinary class is not included in this study). Noticeably, it is more difficult for knowledge from other classes to flow into Economics & Business (ID: 6), suggesting that this is a more independent class that is mainly dependent on the knowledge it creates by itself. Knowledge from biomedicine-, chemistry-, and physics-related classes can be accessed more easily by other classes, whereas it is more difficult for the class of Social Sciences, General to export their knowledge.

| | 1 | 2 | 3 | 4 | 5 | 6 | 7 | 8 | 9 | 10 | 11 | 12 | 13 | 14 | 15 | 16 | 17 | 18 | 19 | 20 | 21 |
|---|---|---|---|---|---|---|---|---|---|---|---|---|---|---|---|---|---|---|---|---|---|
| 1 | 3.75 | 4.22 | 4.77 | 5.37 | 7.20 | 9.28 | 6.46 | 4.37 | 4.98 | 4.20 | 5.48 | 6.80 | 4.00 | 4.70 | 4.70 | 4.20 | 6.69 | 4.77 | 6.16 | 7.30 | 7.40 |
| 2 | 3.91 | 3.64 | 4.19 | 4.47 | 6.41 | 8.32 | 6.01 | 4.35 | 5.43 | 3.44 | 4.73 | 6.17 | 3.41 | 3.69 | 3.72 | 3.33 | 6.09 | 4.48 | 5.22 | 6.50 | 7.00 |
| 3 | 4.43 | 4.37 | 3.13 | 5.34 | 5.58 | 7.04 | 4.77 | 4.13 | 4.68 | 4.17 | 3.70 | 5.67 | 4.14 | 4.67 | 4.67 | 3.53 | 4.55 | 4.93 | 6.11 | 6.60 | 5.17 |
| 4 | 5.06 | 4.60 | 5.22 | 3.97 | 7.51 | 9.78 | 7.02 | 5.60 | 6.56 | 4.00 | 5.84 | 7.27 | 4.32 | 4.82 | 4.05 | 3.99 | 7.03 | 5.69 | 5.36 | 6.72 | 7.98 |
| 5 | 7.10 | 6.72 | 5.35 | 7.65 | 2.83 | 6.21 | 4.93 | 6.67 | 6.80 | 6.50 | 5.08 | 5.00 | 6.50 | 7.00 | 7.00 | 5.83 | 4.67 | 7.55 | 8.27 | 8.06 | 5.67 |
| 6 | 6.68 | 6.85 | 5.44 | 7.95 | 6.29 | 2.79 | 5.54 | 4.88 | 5.51 | 7.13 | 5.80 | 4.94 | 7.13 | 7.25 | 6.88 | 5.71 | 6.21 | 6.45 | 7.63 | 5.38 | 5.38 |
| 7 | 6.34 | 6.11 | 4.70 | 7.06 | 5.25 | 6.65 | 5.13 | 5.14 | 5.54 | 6.00 | 4.76 | 5.83 | 6.01 | 6.41 | 6.34 | 5.16 | 4.86 | 6.37 | 7.67 | 7.42 | 5.08 |
| 8 | 4.70 | 4.74 | 4.14 | 5.93 | 6.83 | 7.88 | 5.29 | 2.87 | 3.71 | 5.00 | 4.70 | 6.50 | 5.00 | 5.08 | 4.67 | 4.06 | 5.52 | 4.08 | 6.67 | 6.38 | 3.83 |
| 9 | 5.87 | 6.16 | 4.77 | 7.34 | 7.09 | 7.15 | 5.67 | 3.71 | 3.55 | 6.55 | 5.32 | 6.91 | 6.55 | 6.59 | 6.14 | 4.97 | 5.23 | 5.34 | 8.16 | 6.90 | 3.27 |
| 10 | 3.60 | 3.22 | 4.00 | 3.03 | 6.33 | 8.88 | 6.04 | 4.50 | 5.64 | 4.89 | 4.70 | 6.00 | 2.00 | 3.50 | 3.50 | 2.67 | 6.11 | 4.15 | 5.00 | 6.34 | 7.00 |
| 11 | 5.56 | 5.19 | 3.86 | 6.16 | 5.25 | 7.18 | 4.88 | 4.83 | 5.39 | 5.00 | 3.79 | 6.15 | 5.00 | 5.45 | 5.40 | 4.37 | 4.53 | 5.78 | 6.97 | 7.33 | 5.10 |
| 12 | 6.30 | 6.17 | 5.63 | 7.06 | 5.00 | 3.88 | 5.33 | 5.88 | 6.41 | 6.00 | 6.05 | 4.50 | 6.00 | 6.50 | 6.50 | 5.67 | 4.61 | 6.79 | 7.43 | 6.40 | 5.00 |
| 13 | 3.87 | 3.82 | 4.67 | 4.33 | 7.00 | 9.54 | 6.60 | 4.83 | 5.70 | 2.33 | 5.37 | 6.67 | 2.83 | 4.17 | 4.17 | 3.56 | 6.78 | 4.46 | 5.67 | 7.16 | 7.67 |
| 14 | 3.85 | 3.11 | 4.25 | 4.15 | 6.58 | 9.13 | 6.23 | 4.46 | 5.66 | 3.25 | 4.90 | 6.25 | 3.25 | 2.92 | 3.75 | 3.25 | 6.31 | 4.42 | 5.21 | 6.56 | 7.25 |
| 15 | 4.10 | 3.61 | 4.50 | 3.90 | 6.83 | 9.38 | 6.54 | 5.00 | 6.14 | 3.50 | 5.20 | 6.50 | 3.50 | 4.00 | 2.00 | 3.17 | 6.39 | 4.81 | 3.57 | 6.33 | 7.50 |
| 16 | 3.93 | 3.56 | 3.61 | 4.36 | 6.11 | 8.21 | 5.57 | 4.00 | 5.00 | 3.33 | 4.47 | 5.83 | 3.33 | 3.83 | 3.17 | 2.33 | 5.56 | 4.41 | 4.67 | 6.25 | 5.67 |
| 17 | 6.56 | 6.43 | 4.70 | 7.16 | 4.93 | 4.88 | 4.89 | 5.56 | 5.62 | 6.22 | 4.70 | 5.44 | 6.22 | 6.72 | 6.50 | 5.59 | 3.35 | 6.66 | 7.60 | 6.97 | 3.22 |
| 18 | 4.55 | 4.51 | 4.74 | 5.48 | 7.18 | 8.88 | 6.23 | 3.64 | 4.63 | 4.31 | 5.28 | 6.77 | 4.28 | 4.81 | 4.39 | 4.18 | 6.41 | 4.12 | 6.23 | 7.05 | 6.00 |
| 19 | 5.86 | 5.38 | 6.07 | 5.34 | 7.71 | 9.09 | 7.57 | 6.50 | 7.54 | 5.21 | 6.69 | 7.68 | 5.24 | 5.73 | 4.00 | 4.55 | 7.67 | 6.56 | 3.81 | 6.35 | 8.64 |
| 20 | 6.37 | 6.20 | 6.05 | 6.30 | 7.93 | 7.14 | 7.01 | 5.43 | 6.23 | 6.03 | 6.59 | 7.25 | 6.16 | 6.49 | 5.81 | 5.24 | 7.26 | 6.47 | 6.12 | 6.15 | 6.86 |
| 21 | 7.20 | 7.22 | 5.25 | 8.10 | 6.00 | 4.88 | 5.56 | 6.00 | 3.46 | 7.00 | 5.40 | 6.00 | 7.00 | 7.50 | 7.50 | 6.33 | 3.22 | 6.92 | 8.43 | 7.57 | 6.56 |

Figure 7. Heat map on shortest path length for ESI classes

In Figure 8, Economics & Business, Engineering, Psychiatry/Psychology, and Social Science, General are more different to/from knowledge of other classes. These classes depend more on their own knowledge: they import and export less knowledge from other classes and thus have a longer citation distance from other classes. Biomedicine-, chemistry-, and physics-related classes, on the other hand, are connected more closely by other classes.

| | 1 | 2 | 3 | 4 | 5 | 6 | 7 | 8 | 9 | 10 | 11 | 12 | 13 | 14 | 15 | 16 | 17 | 18 | 19 | 20 | 21 |
|---|---|---|---|---|---|---|---|---|---|---|---|---|---|---|---|---|---|---|---|---|---|
| 1 | 0.10 | 0.13 | 0.09 | 0.17 | 0.17 | 0.48 | 0.35 | 0.11 | 0.19 | 0.08 | 0.20 | 0.17 | 0.10 | 0.10 | 0.09 | 0.09 | 0.13 | 0.14 | 0.64 | 0.64 | 0.13 |
| 2 | 0.15 | 0.16 | 0.12 | 0.19 | 0.19 | 0.50 | 0.38 | 0.14 | 0.22 | 0.11 | 0.22 | 0.19 | 0.13 | 0.12 | 0.11 | 0.11 | 0.15 | 0.17 | 0.66 | 0.66 | 0.16 |
| 3 | 0.08 | 0.09 | 0.04 | 0.13 | 0.10 | 0.41 | 0.29 | 0.07 | 0.14 | 0.05 | 0.14 | 0.12 | 0.07 | 0.06 | 0.05 | 0.05 | 0.07 | 0.10 | 0.60 | 0.59 | 0.07 |
| 4 | 0.16 | 0.17 | 0.13 | 0.19 | 0.20 | 0.51 | 0.39 | 0.16 | 0.24 | 0.12 | 0.24 | 0.21 | 0.14 | 0.14 | 0.12 | 0.12 | 0.17 | 0.18 | 0.67 | 0.66 | 0.17 |
| 5 | 0.17 | 0.18 | 0.11 | 0.21 | 0.14 | 0.46 | 0.35 | 0.15 | 0.22 | 0.13 | 0.21 | 0.17 | 0.15 | 0.15 | 0.13 | 0.13 | 0.13 | 0.18 | 0.68 | 0.66 | 0.14 |
| 6 | 0.55 | 0.56 | 0.50 | 0.60 | 0.55 | 0.51 | 0.72 | 0.49 | 0.57 | 0.52 | 0.60 | 0.54 | 0.53 | 0.53 | 0.52 | 0.50 | 0.53 | 0.54 | 1.01 | 0.90 | 0.52 |
| 7 | 0.33 | 0.34 | 0.28 | 0.38 | 0.33 | 0.63 | 0.52 | 0.30 | 0.37 | 0.30 | 0.38 | 0.35 | 0.31 | 0.31 | 0.30 | 0.29 | 0.30 | 0.34 | 0.85 | 0.83 | 0.30 |
| 8 | 0.11 | 0.12 | 0.07 | 0.16 | 0.14 | 0.44 | 0.31 | 0.05 | 0.13 | 0.08 | 0.17 | 0.15 | 0.09 | 0.09 | 0.08 | 0.06 | 0.10 | 0.09 | 0.63 | 0.61 | 0.09 |
| 9 | 0.18 | 0.19 | 0.13 | 0.23 | 0.20 | 0.50 | 0.37 | 0.12 | 0.17 | 0.15 | 0.23 | 0.21 | 0.17 | 0.17 | 0.15 | 0.14 | 0.16 | 0.17 | 0.70 | 0.67 | 0.12 |
| 10 | 0.07 | 0.08 | 0.04 | 0.10 | 0.11 | 0.42 | 0.30 | 0.07 | 0.15 | 0.23 | 0.15 | 0.11 | 0.03 | 0.04 | 0.03 | 0.03 | 0.08 | 0.09 | 0.58 | 0.58 | 0.08 |
| 11 | 0.22 | 0.23 | 0.17 | 0.27 | 0.23 | 0.54 | 0.42 | 0.20 | 0.27 | 0.18 | 0.27 | 0.25 | 0.20 | 0.20 | 0.18 | 0.18 | 0.19 | 0.23 | 0.74 | 0.72 | 0.19 |
| 12 | 0.17 | 0.17 | 0.12 | 0.21 | 0.16 | 0.35 | 0.36 | 0.15 | 0.23 | 0.13 | 0.23 | 0.13 | 0.14 | 0.14 | 0.13 | 0.13 | 0.13 | 0.18 | 0.67 | 0.62 | 0.13 |
| 13 | 0.13 | 0.14 | 0.10 | 0.17 | 0.17 | 0.49 | 0.36 | 0.13 | 0.21 | 0.08 | 0.21 | 0.18 | 0.10 | 0.11 | 0.09 | 0.09 | 0.14 | 0.14 | 0.65 | 0.64 | 0.14 |
| 14 | 0.10 | 0.11 | 0.08 | 0.15 | 0.15 | 0.46 | 0.33 | 0.10 | 0.19 | 0.06 | 0.18 | 0.15 | 0.08 | 0.08 | 0.06 | 0.06 | 0.11 | 0.12 | 0.62 | 0.62 | 0.11 |
| 15 | 0.07 | 0.08 | 0.04 | 0.11 | 0.11 | 0.43 | 0.30 | 0.07 | 0.16 | 0.03 | 0.15 | 0.12 | 0.05 | 0.05 | 0.01 | 0.03 | 0.08 | 0.09 | 0.57 | 0.58 | 0.08 |
| 16 | 0.08 | 0.09 | 0.05 | 0.12 | 0.12 | 0.43 | 0.30 | 0.06 | 0.14 | 0.04 | 0.15 | 0.12 | 0.06 | 0.06 | 0.03 | 0.03 | 0.08 | 0.09 | 0.59 | 0.59 | 0.08 |
| 17 | 0.12 | 0.13 | 0.06 | 0.16 | 0.12 | 0.40 | 0.31 | 0.10 | 0.17 | 0.08 | 0.16 | 0.13 | 0.10 | 0.10 | 0.08 | 0.08 | 0.07 | 0.13 | 0.63 | 0.61 | 0.05 |
| 18 | 0.15 | 0.16 | 0.12 | 0.20 | 0.19 | 0.50 | 0.37 | 0.11 | 0.20 | 0.11 | 0.22 | 0.20 | 0.13 | 0.13 | 0.11 | 0.11 | 0.16 | 0.14 | 0.67 | 0.66 | 0.15 |
| 19 | 0.54 | 0.54 | 0.51 | 0.57 | 0.57 | 0.83 | 0.76 | 0.53 | 0.61 | 0.50 | 0.61 | 0.57 | 0.51 | 0.51 | 0.48 | 0.49 | 0.54 | 0.55 | 1.00 | 1.00 | 0.54 |
| 20 | 0.83 | 0.84 | 0.79 | 0.86 | 0.86 | 1.04 | 1.04 | 0.79 | 0.87 | 0.79 | 0.89 | 0.86 | 0.81 | 0.81 | 0.79 | 0.79 | 0.82 | 0.83 | 1.30 | 1.24 | 0.82 |
| 21 | 0.11 | 0.12 | 0.06 | 0.16 | 0.12 | 0.39 | 0.31 | 0.09 | 0.14 | 0.08 | 0.16 | 0.12 | 0.09 | 0.09 | 0.08 | 0.08 | 0.05 | 0.13 | 0.63 | 0.61 | 0.26 |

Figure 8. Heat map on shortest path weight (*1E-3) for ESI classes

As shown in Table 4, knowledge from Pharmacology & Toxicology and Immunology flows to other classes more easily; knowledge from other classes flows into Chemistry and Pharmacology & Toxicology more easily. In terms of their citation distances, knowledge from Neuroscience & Behavior and Pharmacology & Toxicology has a shorter citation distance to other classes. The results may suggest that these domains are comparatively more porous and permeable. Knowledge of Psychiatry/Psychology and Social Sciences, General are more isolated from other classes, which may suggest that they are more self-contained.

Table 4. Average shortest path length and average shortest path weight for 21 ESI classes

| ID | ESI classes | Average shortest path length | | Average shortest path weight | |
|---|---|---|---|---|---|
| | | Source | Destination | Source | Destination |
| 1 | Agricultural Sciences | 5.89 | 5.52 | 2.90E-04 | 3.16E-04 |
| 2 | Biology & Biochemistry | 5.26 | 5.31 | 3.15E-04 | 3.24E-04 |
| 3 | Chemistry | 5.16 | 4.99 | 2.44E-04 | 2.77E-04 |
| 4 | Clinical Medicine | 5.85 | 5.77 | 3.24E-04 | 3.54E-04 |
| 5 | Computer Science | 6.67 | 6.65 | 3.17E-04 | 3.40E-04 |
| 6 | Economics & Business | 6.23 | 7.69 | 6.56E-04 | 6.08E-04 |
| 7 | Engineering | 6.24 | 6.13 | 4.82E-04 | 5.24E-04 |
| 8 | Environment/Ecology | 5.47 | 5.06 | 2.63E-04 | 2.92E-04 |
| 9 | Geosciences | 6.23 | 5.74 | 3.31E-04 | 3.68E-04 |
| 10 | Immunology | 4.90 | 5.11 | 2.36E-04 | 2.77E-04 |
| 11 | Materials Science | 5.76 | 5.43 | 3.75E-04 | 3.77E-04 |
| 12 | Mathematics | 6.15 | 6.54 | 3.07E-04 | 3.46E-04 |
| 13 | Microbiology | 5.63 | 5.18 | 2.98E-04 | 2.95E-04 |
| 14 | Molecular Biology & Genetics | 5.27 | 5.60 | 2.72E-04 | 2.94E-04 |
| 15 | Neuroscience & Behavior | 5.27 | 5.14 | 2.36E-04 | 2.76E-04 |
| 16 | Pharmacology & Toxicology | 4.98 | 4.56 | 2.43E-04 | 2.74E-04 |
| 17 | Physics | 6.07 | 6.06 | 2.67E-04 | 3.06E-04 |
| 18 | Plant & Animal Science | 5.76 | 5.76 | 3.13E-04 | 3.25E-04 |
| 19 | Psychiatry/Psychology | 6.33 | 6.28 | 6.85E-04 | 8.14E-04 |
| 20 | Social Sciences, General | 6.44 | 6.72 | 9.63E-04 | 7.91E-04 |
| 21 | Space Science | 6.58 | 6.25 | 2.61E-04 | 3.02E-04 |

### Results on top divisions of knowledge

Subject categories can further be aggregated into sciences and social sciences as a whole: there are 170 science disciplines and 51 social science disciplines. The average shortest path length within science disciplines is 5.49, which is shorter than the average shortest path length within social science disciplines (6.13). Figure 9 also suggests that it is easier for social science knowledge to flow into science disciplines than for science knowledge to flow into social science disciplines. The results may be attributed to the fact that quite a few social sciences are more independent in that they primarily cite publications within their own disciplines, resulting in a longer knowledge path from science to social science. Distance-wise, science disciplines are related closely via citations (average shortest path weight is 0.24*1E-3), and social science disciplines are more disunified (average shortest path weight is 1.16*1E-3).

| | Science | Social Science |
|---|---|---|
| Science | 5.49 | 6.99 |
| Social Science | 6.53 | 6.13 |

*1E-3

| | Science | Social Science |
|---|---|---|
| Science | 0.24 | 0.71 |
| Social Science | 0.78 | 1.16 |

Figure 9. Average shortest path length and weight for top divisions of knowledge

## Discussion

### Patterns of knowledge paths

Figure 10 summarizes eight different knowledge path types between science (S) and social science (SS). The first number of each type shows the number of paths that fall into this type, the second number shows its percentage in relation to all knowledge paths, and the third number shows its percentage in relation to the block (i.e. S->S, S->SS, SS->S, or SS->SS).

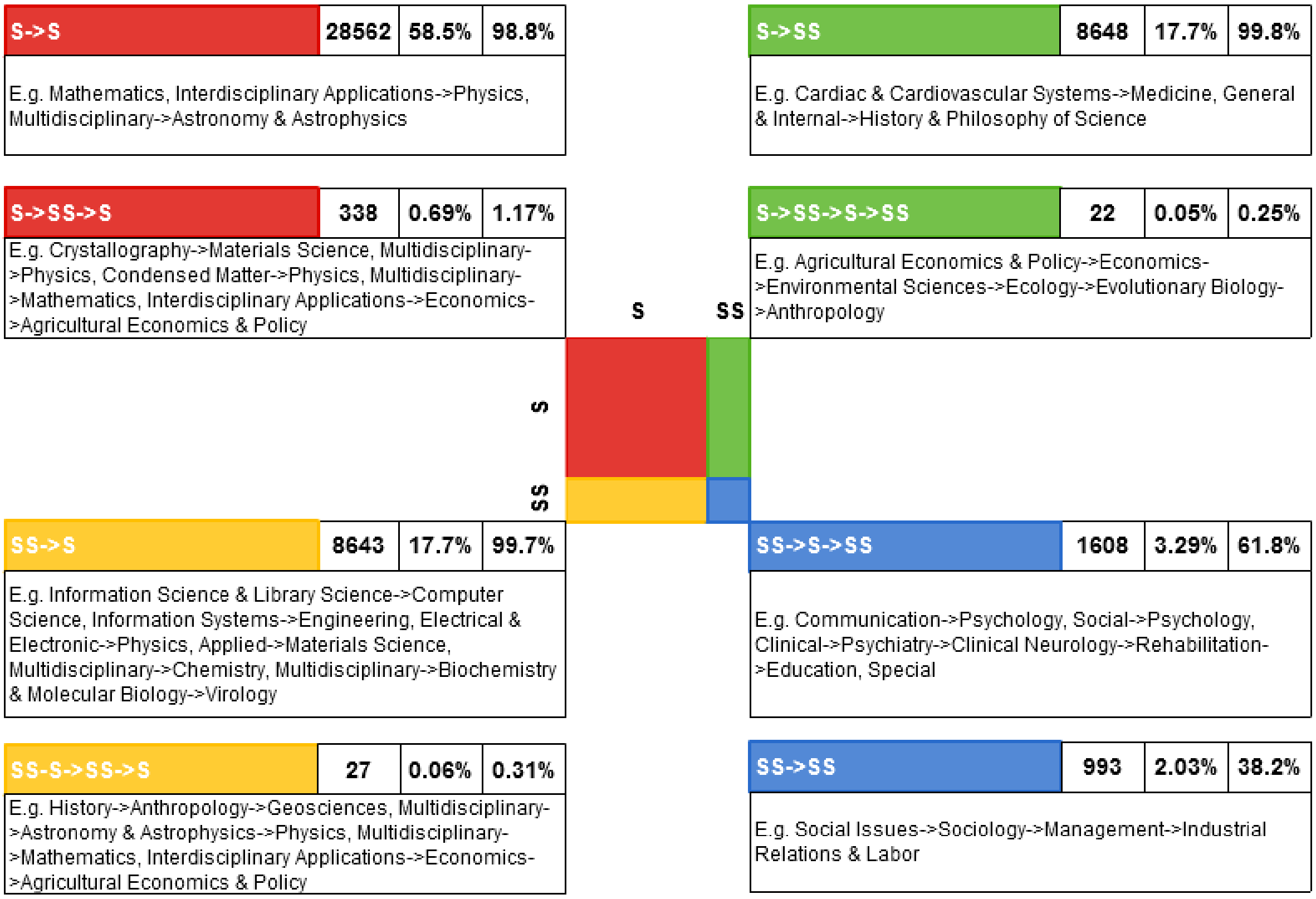

Figure 10. Knowledge path types

Knowledge flows from one science discipline to another science discipline mainly stay within science disciplines, and only 1 percent of the knowledge paths need one or a few social science disciplines as their intermediate (S->SS->S). Nevertheless, from one social science discipline to another social science discipline, up to 62 percent of the knowledge paths need one or a few science disciplines as the intermediate (SS->S->SS), suggesting the indispensable role of science disciplines in transferring social science knowledge. This confirms the disunity of social science disciplines as conceptually studied by Suppes (1984) and Habermas (1988). The knowledge flow between a science discipline and a social science discipline is more direct, with a limited number of detours.

## Knowledge flow maps

Knowledge flow networks are visualized herein, where each node represents a subject category and each link represents a knowledge flow. The size of the node corresponds to their PageRank scores and the size of the link corresponds to the widths of their knowledge flow. The calculation of nodes' PageRank scores and visualizations of the knowledge flow networks are based on an online service Map Equation (Rosvall & Bergstrom, 2008).

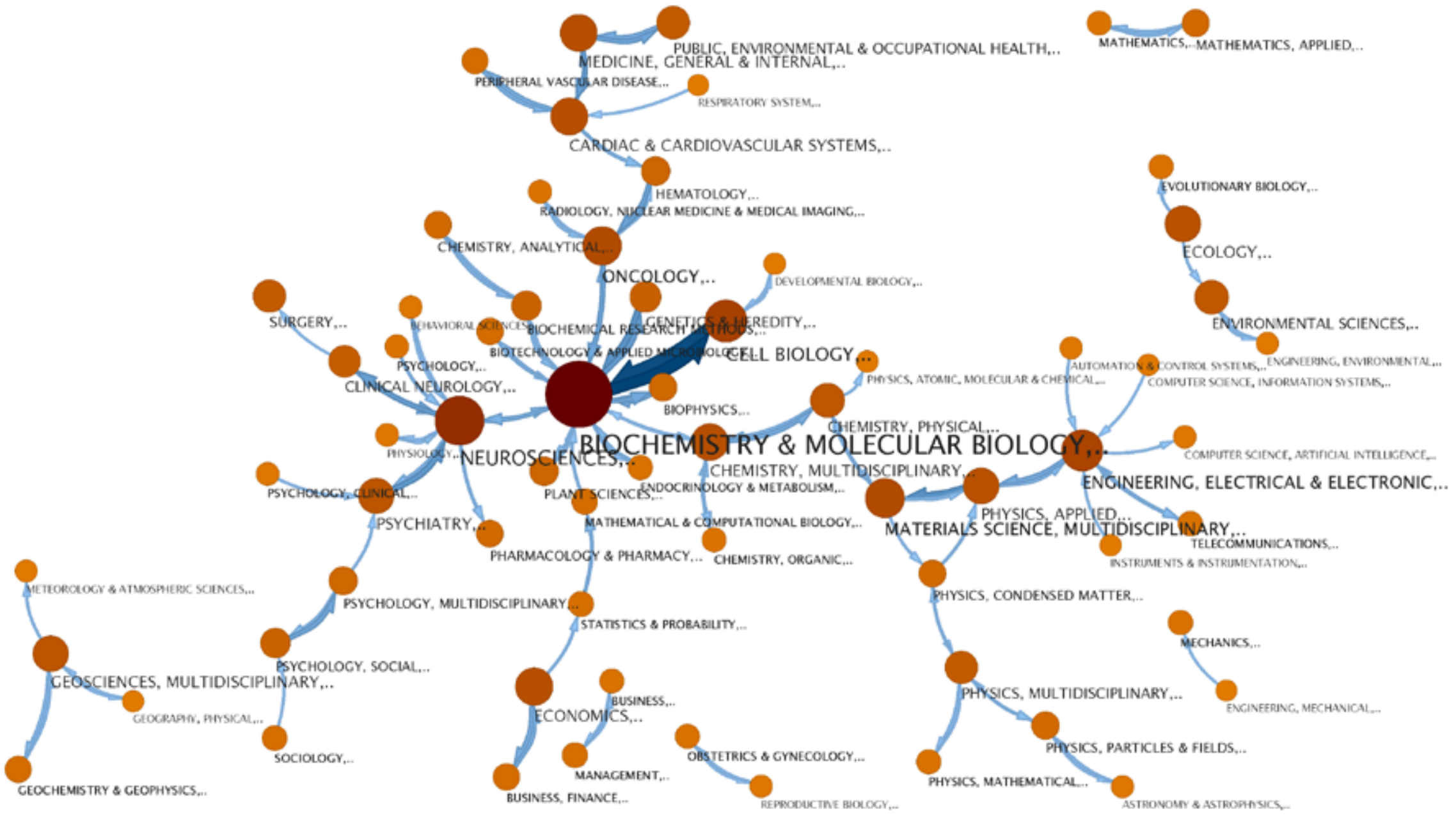

Figure 11. Knowledge flow map for all science and social science disciplines

In Figure 11, several knowledge hubs can be found, including (1) Biochemistry & Molecular Biology, (2) Medicine, General & Internal, (3) Economics, (4) Chemistry, Multidisciplinary, (5) Materials Science, Multidisciplinary, (6) Physics, Multidisciplinary, (7) Engineering, Electrical & Electronics, and (8) Psychiatry. Centered by Biochemistry & Molecular Biology, those disciplines form a clear backbone of science.

Major knowledge paths include:

- From Biochemistry & Molecular Biology, to Cell Biology, and to Environmental Biology;
- From Biochemistry & Molecular Biology, to Chemistry, Multidisciplinary, to Materials Science, Multidisciplinary, to Physics, Applied, and to Engineering, Electrical & Electronics;
- From Biochemistry & Molecular Biology, to Oncology, to Cardiac & Cardiovascular Systems, to Medicine, General & Internal, and to Public, Environmental, & Occupational Health;
- From Biochemistry & Molecular Biology, to Mathematical & Computational Biology, to Statistics & Probability, to Economics, and to Business & Finance; and

- From Biochemistry & Molecular Biology, to Neuroscience, to Psychiatry, to Psychology, Multidisciplinary, and to Sociology.

Most of these knowledge paths are bidirectional, for instance, the knowledge path between Biochemistry & Molecular Biology and Cell Biology has a reciprocal flow. Meanwhile, for some knowledge paths, the amount of incoming knowledge is either much smaller or much larger than the amount of outgoing knowledge. These are seen as unidirectional arrows in Figure 11, for instance, from Statistics & Probability to Mathematical & Computational Biology, from Geoscience, Multidisciplinary to Meteorology & Atmosphere Science, and from Cardiac & Cardiovascular Systems to Hematology.

In Figure 11, Psychology and Economics connect knowledge between the sciences and social sciences. The results confirm earlier findings where psychology- and economics-related fields occur more frequently in shortest paths. As most social science fields are not visible in Figure 11, a visualization that only includes social science fields is illustrated in Figure 12.

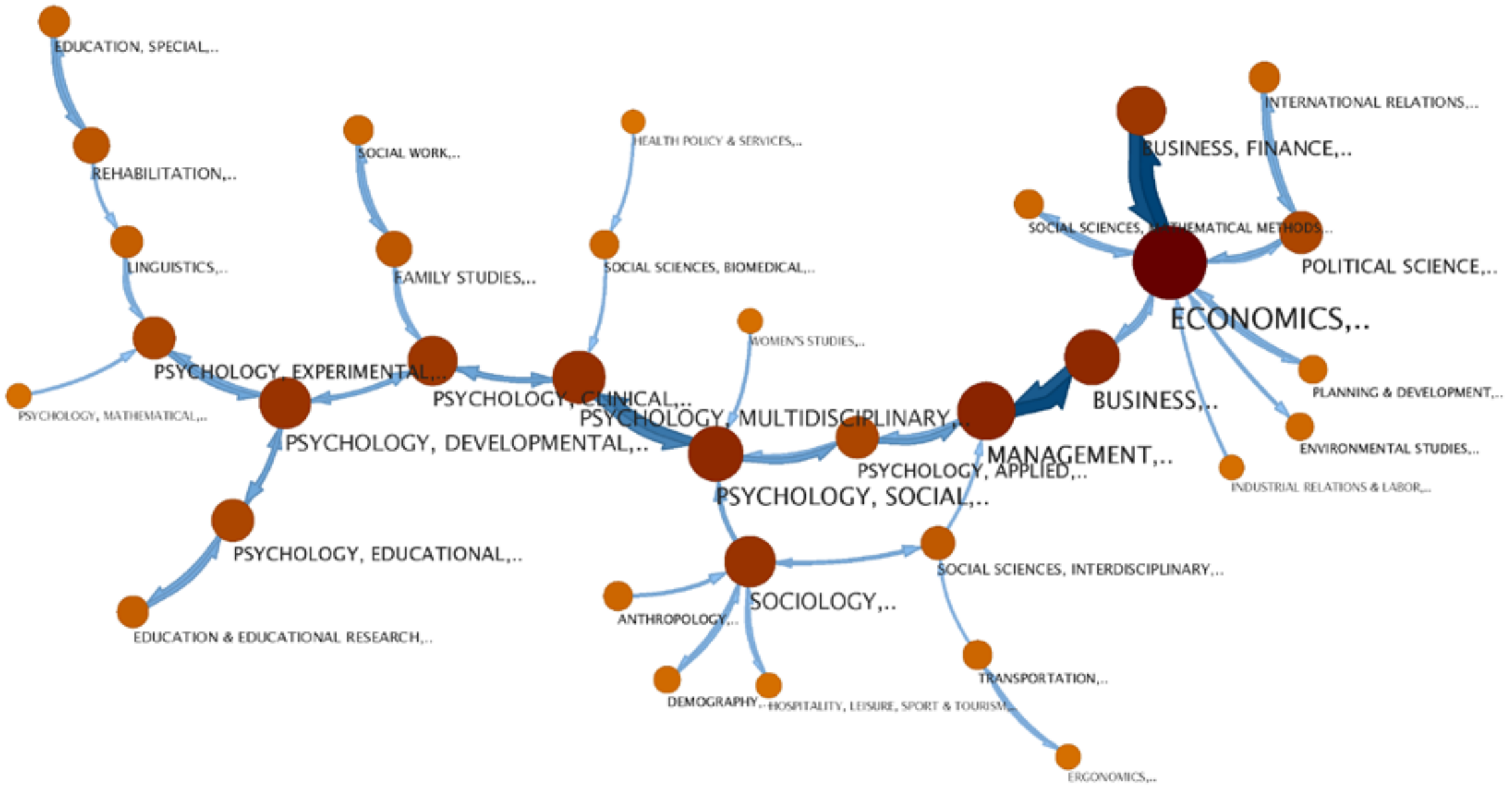

Figure 12. Knowledge flow map for social science disciplines

The “backbone” of social science is evident in Figure 12. The bidirectional knowledge path for social sciences is from (1) Political Science, to (2) Economics, to Business, to (3) Management, to (4) Sociology, to (5) Psychology, Social, to (6) Psychology, Developmental, to (7) Education & Educational Research, and finally, to (8) Linguistics. The knowledge flow of other social science disciplines is facilitated through this backbone knowledge path.

**Unified science or disunified science?**

The goal of this study is not to provide an absolute answer on whether science is unified or disunified, but to provide empirical analyses on the connectivity of different science and social science disciplines through the application of knowledge paths. ISI has incorporated the notion of unity of science by allowing a journal to be assigned into multiple subject categories. This may not be a perfect solution, but according to an interview with Garfield (Garfield & Stock, 2002), the *Science Citation Index* is the "empirical solution" to the unity of science movement (Neurath, 1996). Subject categories with a higher number of multi-assigned journals (e.g., Biochemistry & Molecular Biology, Chemistry, Multidisciplinary, and Social Science, Interdisciplinary) exemplify "unity of science", and thus, empirically, in this study, they tended to possess shorter average knowledge paths.

Using the ISI data, this study finds that biomedicine-, chemistry-, and physics-related domains can access and be accessed by other domains more easily. The results may imply that these domains have a higher level of unification than other domains. Reasons behind this phenomenon are not conclusive, but can be interpreted from several angles. For instance, it is likely that biomedicine-, chemistry-, and physics-related domains share "a very narrow and homogeneous class of terms of the physical thing-language" (Carnap, 1955, p. 404). The "unity of language" may have facilitated the communication of scientists in these domains and thus have helped them address questions that have common grounds (Habermas, 1988). At the same time, language may also define and shape disciplines, especially for some social science domains, as Hyland (2004) emphasized that "writing…[o]n the contrary, it helps to create those disciplines". Another defining feature that may lead to the unification of science is the "unity of laws" (Carnap, 1955). Even though "there is at present no unity of laws" (p. 403) for the whole science, at the field level, a few fields may share a higher level of unity of laws. For instance, this study shows that knowledge can flow more easily among biomedicine-, chemistry-, and physics-related domains than that from these domains to social science and psychology-related domains. The results may be attributed to the fact that "laws of psychology and social science cannot be derived from those of biology and physics" (p. 403). In addition to the unity of language and unity of laws, in Margenau's (1941) review article on John Dewey's *Unity of Science as a Social Problem*, two aspects on unity of science are espoused: "synthetic unity of the attainments of the individual sciences…[and] unity and universality of scientific attitude" (p. 433). The unity of attainments has not been achieved, according to Margenau (1941), but the unity of scientific attitude is expected to be ubiquitous: "it is the will to inquire, to examine, to discriminate, to draw conclusions only on the basis of evidence" (p. 433).

Historically, there is a tendency towards a unified science (Neurath, 1996). The classic "universitas literarum" (Neurath, 1938) still has significance in current scientific exploration. The emerging field of big data research (e.g., World Economic Forum, 2012) echoes the notion of “departmentalized into special science, and not toward a speculative juxtaposition of an autonomous philosophy and a group of scientific disciplines” (Neurath, 1996, p. 328). A longitudinal study (e.g., Yan, Ding, & Kong, 2013) may prove to be necessary to examine the level of unification and thus providing insights on the evolving character of science.

**The classification of science**

The limitation of the knowledge flow networks is that they only represent the knowledge classification by ISI (Thomason Reuters). ISI assigns journals to subject categories based on journal-to-journal citation patterns and editorial judgment (Pudovkin & Garfield, 2002). According to Leydesdorff and Rafols (2009), subject categories cannot be considered as literary warrant (Chan, 1999) like the Library of Congress Classification. A “multidimensional journal evaluation” may thus alleviate this tension (Haustein, 2012): in addition to journal citations, journal output, content, perception, usage, and management should all be considered. Subject categories may also have coverage limitations (e.g., Meho & Yang, 2007): biomedical related fields are better presented than social science related fields; meanwhile, different fields may have varied community size and citation and/or publishing patterns (Guerrero-Bote et al., 2007), and thus, for impact analysis of paper, authors, or journals, field-level normalization has been proven to be necessary (e.g., Waltman, Van Eck, Van Leeuwen, Visser, & Van Raan, 2010). The focus of this paper, however, is not to evaluate research impact (i.e., to evaluate which subject category has a higher impact), but to examine how knowledge flows among different fields. Therefore, in the context of scientific trading (Yan, Ding, Cronin, & Leydesdorff, 2013), each field is considered as a single trading unit. Yet, different fields vary greatly in size (e.g., number of publications and number of journals), the use of fields as the unit of analysis should be further examined. Because subject categories are used as a proxy to map a variety of fields, the caveat is the likely inconsistencies between the actual status quo of scientific fields and how they are framed into subject categories.

Through journal co-authorship network analysis, Ni, Sugimoto, and Jiang (2013) found that four well defined fields can be identified in JCR subject category of Information Science and Library Science (ISLS): information science, library science, professional studies, and management information science (MIS). In particular, MIS journals possessed a loose citation linkage with other journals in the subject category. The authors recommended that MIS journals may be removed from the ISLS subject category and may have a standalone subject category. Using the same data set as the current study, Yan and colleagues (2013) identified several subject categories that have very low self-dependence, for example, (1) Psychology, Biological, (2) Social Science, Mathematical Methods, (3) Medicine, Research & Experimental, (4) Biology, and (5) Anatomy & Morphology. These subject categories primarily cited publications in other subject categories but not their own publications. It is therefore counterintuitive to keep them as independent subject categories. In addition to issues pertaining to subject categories, their higher level classes, Essential Science Indicator, could also be proven. Out of the 21 classes under study, there are only three social science related classes. The class Social Science, General comprises around 50 subject categories; at the same time, the class Immunology only comprises one subject category – Immunology.

This inequality of visibility needs to be explored further as to determine whether ISI classification is capable of presenting the actual visibilities of a variety of fields in academics or if there is a discrepancy between the actual visibility and the visibility captured by ISI. As the visibility of a field can be exemplified in many forms, a possible solution is to incorporate different sources of data, in addition to publication and citation data. For instance, books are an important channel in scholarly communication, as Garfield and Stock (2002) pointed out in *Science Citation Index*, that 15% cited references are non-journal items with a majority being books, and in *Social Science Citation Index*, that 50% of cited references are books. Textbooks, usually not well examined through bibliometric lens, are seen as an important indicator to analyze how "normal science" is codified and presented, as Kuhn (1970) argued that "…mainly from the study of finished scientific achievements as these are recorded in the classics and, more recently, in the textbooks from which each new scientific generation learns to practice its trade." (p. 1). Thus, books/textbooks, number of enrolled students, faculty members, courses provided and so on should also be considered in assessing disciplinary visibilities.

## Conclusion

In this article, a knowledge flow network is constructed to study how knowledge is disseminated among various disciplines. In order to systematically study such knowledge transfer, a knowledge hierarchy is proposed that includes top divisions of knowledge, ESI classes, JCR subject categories, journals, and papers. Knowledge flow in the hierarchies of sciences/social sciences, ESI classes, and JCR subjects is investigated herein.

Quantitative indicators, including shortest path length, shortest path weight, and occurrence in shortest path, are proposed and applied. Based on an investigating of subject categories of Journal Citation Report, it is found that social science domains tend to be more independent and thus differ from other domains as measured by citation distance. Consequently, it is more difficult for knowledge from other domains to flow into social science domains. Knowledge from science domains, such as biomedicine, chemistry, and physics, can be accessed more easily by other domains. It is also found that social science domains are more disunified than science domains, as up to three-fifths of the knowledge paths within social sciences need at least one science discipline to serve as their intermediate to connect two social science disciplines.

This paper examines patterns of knowledge dissemination and empirically investigates the issue of disciplinarities and interdisciplinarities. Future studies in this direction will benefit from exploring the dynamic aspects of knowledge dissemination as well as adding semantics to knowledge flow representations. Such contextualization will be particularly valuable for scholars when navigating among concrete research concepts, theories, or methods, and when investigating the provenance and inheritance of various research entities.

## Acknowledgements

The author would like to thank Dr. Loet Leydesdorff of the University of Amsterdam for providing access to the data set. The author would also like to thank Dr. Ying Ding, Dr. Katy Börner, and Dr. Cassidy Sugimoto of Indiana University for their comments on an earlier draft and an informal presentation of this paper.